\documentclass[amsmath,amssymb,aps,prl,twocolumn]{revtex4-2}
\usepackage{graphicx}
\usepackage{amsmath}
\usepackage{hyperref}
\usepackage{blkarray}
\usepackage{physics}
\usepackage{bbold}
\usepackage[dvipsnames]{xcolor}
\usepackage{tikz}
\usetikzlibrary{plotmarks}
\usepackage{standalone}
\usepackage[caption=false]{subfig}

\newcommand{\Psibar}{\overline{\Psi}}

\newcommand{\psip}{\psi_{+}}
\newcommand{\psim}{\psi_{-}}
\newcommand{\Psiplus}{\Psi^{(+)}}
\newcommand{\Psiminus}{\Psi^{(-)}}

\renewcommand{\section}[1]{\emph{#1}.---}
\renewcommand{\subsection}[1]{\relax}

\graphicspath{{./}}

\begin{document}

\title{Entanglement and magic on the light-front}
\author{Sam Alterman}
\email{sam.alterman@tufts.edu}
\author{Peter J. Love}
\email{peter.love@tufts.edu}
\affiliation{Department of Physics and Astronomy, Tufts University, Medford, Massachusetts 02155, USA}
\date{July 14, 2025}
\begin{abstract}
    In the light-front (LF) formulation of quantum field theory (QFT), physics is formulated from the perspective of a massless observer necessarily traveling at the speed of light. The LF formulation provides an alternative computational approach to lattice gauge theory, and has recently been investigated as a future application of quantum computers. A natural question is how quantum resources such as entanglement and contextuality amongst physical qubits in the laboratory are utilized in LF simulations of QFTs.  We use the (1+1)D transverse-field Ising model to explore this question. We derive the LF energy operator that generates the LF dynamics of the system, which is distinct from the instant-form (IF) Hamiltonian. We find that while the eigenstates of the IF Hamiltonian exhibit pairwise entanglement between positive and negative momenta in IF momentum-space, the eigenstates of the LF Hamiltonian are separable in LF momentum-space. We then calculate the momentum-space magic of the IF-momentum-space ground state and show that it always requires more magic to prepare than the LF-momentum-space ground state. At the quantum critical point, corresponding to a massless free fermion,  both LF and IF ground states are stabilizers, but the LF ground state is separable in LF momentum-space while the IF ground state is a product of maximally entangled pairs in IF momentum-space. These results show that quantum  resources such as entanglement and magic are utilized differently by quantum simulations formulated in LF and IF, and that the simplicity of the LF ground state results in fewer required quantum resources.
\end{abstract}

\maketitle
\newpage
\section{Introduction}
Quantum field theory (QFT) is a natural application for which quantum computing may perform simulations beyond the capabilities of classical computers \cite{BauerHEP,BauerParticlesForces,Jordan2012,Jordan2014,Jordan2018}. Recently, the question of how quantum resources such as entanglement, contextuality and magic can be most efficiently utilized in quantum simulations of QFT has attracted much attention~\cite{Klco2022,Calabrese2004,Buividovich2009,Casini2014,Klco2023a,Klco2023b}. Most approaches to quantum simulations of QFT use the instant form (IF) representation of QFTs in which all observables lie on a spacelike hypersurface. IF coordinates $x^0$ and $x^1$ describe Minkowskian spacetime as seen by a massive observer in an inertial reference frame with the (1+1)D metric $g_{00}=-g_{11}=1,$ $g_{10}=g_{01}=0$. However, Dirac \cite{DiracLF} showed that the IF is not the only valid formulation of relativistic quantum mechanics. Light-front (LF) coordinates describe the perspective of a massless observer. LF coordinates in (1+1)D are defined as $x^{\pm}=x^0\pm x^{1}$, with $x^+$ playing the role of light-front ``time'' and $x^-$ playing the role of light-front ``space''. The metric is $g^{+-}=g^{-+}=2,\quad g^{++}=g^{--}=0$~\cite{DiracLF,Pauli1985-1,Pauli1985-2}. It was first realized by Wilson \cite{Wilson1990} that the LF formulation results in representations of QFTs that show a remarkable similarity to well-studied problems in quantum chemistry, and LF simulations of QFTs have recently been investigated as an approach for efficiently simulating quantum field theory on a quantum computer \cite{Kreschuk2021a,Kreshchuk2021,Kreshchuk2022,Kreschuk2023,CarterGPDF,Wu2024,Kreshchuk2024, Lobe}. 

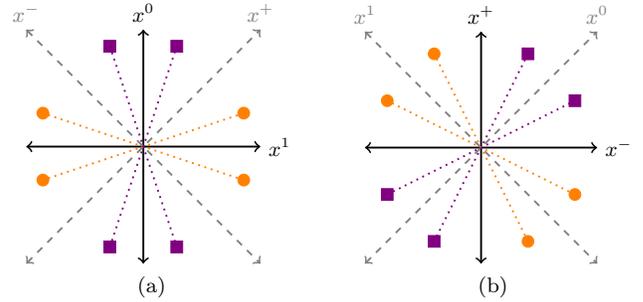
\begin{figure}[ht!]
    \subfloat[\label{fig:ifspacetimediagram}]{
    \begin{tikzpicture}
            \draw[<->,black, thick](0,-1.75) -- (0,1.75) node[anchor=south] {$x^0$};
            \draw[<->,black, thick](-1.75,0) -- (1.75,0) node[anchor=west] {$x^1$};
            \draw[<->,gray, thick,dashed](-1.75,-1.75) -- (1.75,1.75) node[anchor=south]{$x^+$};
            \draw[<->,gray, thick,dashed](-1.75,1.75) node[anchor=south]{$x^-$} -- (1.75,-1.75);
            
            \draw[orange, thick, dotted](1.5,0.5) -- (-1.5,-0.5);
            %\filldraw[red] (1.5,0.5) circle (2pt); %node[anchor=west]{B};
            %\filldraw[red] (-1.5,-0.5) circle (2pt);
            \draw[orange, thick, dotted](1.5,-0.5) -- (-1.5,0.5);
            %\filldraw[red] (1.5,-0.5) circle (2pt);% node[anchor=west]{B};
            %\filldraw[red] (-1.5,0.5) circle (2pt);
            \draw[orange] plot [only marks,mark=*, mark size =2.5pt] coordinates{(1.5,0.5)(-1.5,-0.5)(1.5,-0.5)(-1.5,0.5)};

            \draw[violet,thick,dotted](-0.5,1.5) -- (0.5,-1.5);
            %\filldraw[blue](-0.55,1.45) rectangle (-0.45,1.55);% node[anchor=south]{K-S};
            %\filldraw[blue](0.45,-1.45) rectangle (0.55,-1.55);
            \draw[violet,thick,dotted](-0.5,-1.5) -- (0.5,1.5);
            %\filldraw[blue](0.55,1.45) rectangle (0.45,1.55);% node[anchor=south]{K-S};
            %\filldraw[blue](-0.45,-1.45) rectangle (-0.55,-1.55);
            \draw[violet] plot [only marks,mark=square*,mark size=2.5pt] coordinates {(-0.5,1.5)(0.5,-1.5)(-0.5,-1.5)(0.5,1.5)};
\end{tikzpicture}
    }
    \hfill
    \subfloat[\label{fig:lfspacetimediagram}]{
    \begin{tikzpicture}
            \draw[<->,black, thick](0,-1.75) -- (0,1.75) node[anchor=south] {$x^+$};
            \draw[<->,black, thick](-1.75,0) -- (1.75,0) node[anchor=west] {$x^-$};
            \draw[<->,gray, thick,dashed](-1.75,-1.75) -- (1.75,1.75) node[anchor=south]{$x^0$};
            \draw[<->,gray, thick,dashed](-1.75,1.75) node[anchor=south]{$x^1$} -- (1.75,-1.75);

            \draw[orange, thick, dotted](1.414,-0.707) -- (-1.414,0.707);
            \draw[orange,thick,dotted](0.707,-1.414) -- (-0.707,1.414);
            \draw[orange] plot [only marks,mark=*, mark size =2.5pt] coordinates{(1.414,-0.707)(-1.414,0.707)(0.707,-1.414)(-0.707,1.414)};

            \draw[violet,thick,dotted](0.707,1.414) -- (-0.707,-1.414);
            %\filldraw[blue](0.657,1.364) rectangle (0.757,1.464);
            %\filldraw[blue](-0.757,-1.464) rectangle (-0.657,-1.364);
            \draw[violet,thick,dotted](-1.414,-0.707) -- (1.414,0.707);
            %\filldraw[blue](-1.464,-0.757) rectangle (-1.364,-0.657);
            %\filldraw[blue] (1.364,0.657) rectangle (1.464, 0.757);
            \draw[violet] plot [only marks,mark=square*,mark size=2.5pt] coordinates {(0.707,1.414)(-0.707,-1.414)(-1.414,-0.707)(1.414,0.707)};
        \end{tikzpicture}
    }
    \caption{(a) Spacetime diagram showing Bell type measurements testing entanglement (orange circles) and Kochen-Specker type measurements testing contextuality (purple squares) in instant-time coordinates. (b) The same measurements as viewed in light-front coordinates. A quantum computer where LF space and time in the simulation correspond to IF space and time in the lab would have Bell type measurements in the lab which correspond to Kochen-Specker type measurements in the theory, and vice versa.}
    \label{fig:spacetimefig}
\end{figure}

A natural question is how quantum resources change between the IF and LF representations of a QFT.  In this work, we consider entanglement, contextuality and magic as resources for quantum computing \cite{Spekkens2008,Howard2014,BermejoVega2017}. Suppose we have a quantum computer which we wish to use to simulate a QFT. Assuming the quantum computer is made of massive components its quantum resources will be defined in the (IF) inertial laboratory frame. Mermin \cite{Mermin1993} notes that entanglement can be conceptualized as emerging when non-contextuality is required by locality. Through this framework, Bell tests of entanglement can be associated with spacelike separations while Kochen-Specker tests of contextuality can be associated with timelike measurements, as depicted in Fig. \ref{fig:ifspacetimediagram} \cite{Bell1964,KS1967}.  This perspective allows us to see that if we set up our simulation such that (IF) physical space and time in the laboratory frame correspond to LF space and time respectively in the simulation, the relationship between quantum resources in the QFT and quantum resources on the physical device is non-trivial. As seen in Fig. \ref{fig:lfspacetimediagram}, the structure of LF coordinates means that (spacelike) entanglement between two qubits in the lab frame could correspond to either entanglement or contextuality in the QFT, and entanglement in the QFT could correspond to either (spacelike) entanglement or (timelike) contextuality in the lab frame. Because the utilization of quantum resources in QFTs is an active area of research, we are motivated to study how resources change on the LF through a toy model which is well understood but still allows us to utilize the LF framework.

\subsection{The Ising model}
The transverse-field Ising model (TFIM) is a well studied system that exhibits fermionic behavior in the vicinity of its quantum critical point at $\lambda=1$~\cite{Dutta2015,Toth2005,Mbeng2020,Mussardo2020,Fradkin2021}. This is the critical point of a quantum phase transition between ferromagnetic and antiferromagnetic behavior, with an associated diverging correlation length $\xi\sim a\vert 1-\lambda\vert^{-1}$ \cite{Dutta2015,LSM1961}. In this work, we consider a TFIM consisting of a linear chain of $N$ spins with spacing $a$ and physical length $L=Na$ oriented along the $x$ axis. We assume $N$ to be even. Each spin is coupled to nearest neighbors and subjected to a transverse magnetic field of relative strength $\lambda$ along the $z$ axis, resulting in a Hamiltonian for the system given (in IF coordinates) by\begin{equation}\label{eq:originalIsingHam}
    H_{IF}=-\frac{1}{2}\sum_{j} X_{j}X_{j+1}-\frac{\lambda}{2}\sum_j Z_j
\end{equation}
where $X_j\ (Z_j)$ is a Pauli matrix applied to the $j^\text{th}$ spin in the chain. We will assume $N$ to be even and sufficiently large as to neglect boundary terms. Using the Jordan-Wigner transformation \cite{JordanWigner} to define fermionic operators
\begin{equation}\label{eq:JWops}
    \begin{split}
        c_j&=\frac{X_j-iY_j}{2}\prod_{j'<j}Z_{j'},
    \end{split}
\end{equation}
the Hamiltonian \eqref{eq:originalIsingHam} is expressed in fermionic form as
\begin{equation}\label{eq:origfermiham}
    \begin{split}
    H_{IF}=\sum_j\bigg[&\frac{\lambda}{2}\qty(2c^\dagger_j c_j-1)\\
    &-\frac{1}{2}\qty(c^\dagger_j-c_j)\qty(c^\dagger_{j+1}+c_{j+1})\bigg]
    \end{split}
\end{equation}
where we have assumed the fermionic spatial periodic boundary condition $c_{N+1}=c_1$ \footnote{We note that this fermionic boundary condition is distinct from the spin periodic boundary condition $X_N=X_1$, which splits the Hamiltonian into two orthogonal spaces based on fermion number parity \cite{Mbeng2020}. However, because we will be taking the large $N$ limit and treating the spectrum as continuous, this nuance can be safely neglected.}.  Assuming $N$ to be sufficiently large as to approximate the momentum $k$ as continuous over $(-\pi/a,\pi/a]$, the Hamiltonian \eqref{eq:origfermiham} can be written in momentum space as \cite{Mussardo2020,Fradkin2021,Mbeng2020}
\begin{equation}\label{eq:origbogoham}
\begin{split}
     H&=\int_{0}^{\pi/a} \mathrm{d}k\ 2\omega_k \qty(\hat\gamma^\dagger_k\hat\gamma_k+\hat\gamma^\dagger_{-k}\hat\gamma_{-k}-1) \\
     \omega_k&=\sqrt{\lambda^2-2\lambda\cos{(ka)}+1}
\end{split}
\end{equation}
with $\hat{O}_k$ denoting that an operator $O$ acts in momentum space and $\hat\gamma_k=\cos{\theta_k}\hat c_k+i\sin{\theta_k} \hat c^\dagger_{-k},\ \tan{2\theta_k}=\frac{\sin(ka)}{\lambda-\cos(ka)}.$ A thorough derivation of \eqref{eq:origbogoham} is given in the Appendix.
We can then see that for $ka\ll1,\ \lambda \approx1$, the spectrum becomes the relativistic dispersion relation of a free fermion $\omega_k=\sqrt{m^2+k^2}$ with mass $m=(1-\lambda)/a$.
In the vicinity of the quantum critical point $\lambda=1$, the mass will be determined by how we take the limits $\lambda\rightarrow1$ and $a\rightarrow0$.

To map the system onto a fermionic field, we define at each point on the chain in IF space $x_j=a j$ a field spinor $\Psi(x_j)=\begin{pmatrix}
        c_j\\c_j^\dagger
    \end{pmatrix}.$ Taking the continuum limit $a\rightarrow0$, when $\lambda\rightarrow 1$ (and thus $\xi\gg a$) the IF dynamics of $\Psi$ are modeled by the free fermion Dirac equation  \cite{Ferrell1973}
\begin{equation}\label{eq:IFdirac}
    (i\gamma^0\partial_0+i\gamma^1\partial_1-m)\Psi=0
\end{equation}
 with $x^0=a t$, $x^1=ja$, mass $m=(1-\lambda)/a$, and gamma matrices $\gamma^0=Z$ and $\gamma^1=iX$ \footnote{We note that the gamma matrices do not act on the spins themselves but instead on the spinor $\Psi(x)$, and thus the Pauli matrices used to define the gamma matrices have an entirely distinct physical meaning from the Pauli matrices used to define the IF Hamiltonian \eqref{eq:originalIsingHam}.}.

\section{The TFIM on the light-front}
\subsection{LF dynamics}
We now consider the TFIM on the LF. The evolution is governed by the Dirac equation, written in terms of LF derivatives as
\begin{equation}\label{eq:LFdirac}
    (i\gamma^+\partial_++i\gamma^-\partial_--m)\Psi=0
\end{equation}
with gamma matrices $\gamma^+=\gamma^0+\gamma^1=Z+iX$ and $\gamma^-=\gamma^0-\gamma^1=Z-iX$. We follow~\cite{Mannheim2021} and define orthogonal projectors $\Lambda^{(\pm)}=\frac{1}{4}\gamma^\mp\gamma^\pm=\frac{1}{2}(1\mp Y)$. These projectors allow us to split $\Psi$ into two components $\Psiplus$ and $\Psiminus$ given by 
\begin{equation}
    \Psi^{(\pm)}=\Lambda^{(\pm)}\Psi=\frac{1}{\sqrt{2}}\begin{pmatrix}1\\\mp i\end{pmatrix}\psi_\pm,\ 
    \psi_\pm=\frac{1}{\sqrt{2}}\qty(c\pm i c^\dagger).
\end{equation}
From \eqref{eq:LFdirac}, we obtain one dynamical equation
\begin{subequations}\label{eq:lffreefermievol}
\begin{equation}
     2i\partial_+\Psiplus=m\gamma^0\Psiminus
 \end{equation}
 which gives the variation of $\Psiplus$ (termed the ``good'' fermion) with LF time $x^+$ and one kinematical equation 
 \begin{equation}
     2i\partial_-\Psiminus=m\gamma^0\Psiplus
 \end{equation}
 which gives the variation of $\Psiminus$ (termed the ``bad'' fermion) with LF space $x^-$ \cite{Pauli1985-1,Mannheim2021}. 
\end{subequations}

The equations of motion for $m\neq0$ \eqref{eq:lffreefermievol} are associated with the LF action
 \begin{equation}\label{eq:lfaction}
    \mathcal{S}=\int\mathrm{d}x^+\mathrm{d}x^-\Psibar\qty(i\gamma^+\partial_++i\gamma^-\partial_--m)\Psi.
 \end{equation}
from which we obtain the LF energy operator \cite{Pauli1985-1,Mannheim2021}
\begin{align}\label{eq:LFpminus}
P^-
&=\frac{m}{2}\int_{-\infty}^\infty \mathrm{d} x^- \psip^\dagger\psim.
\end{align}
We will discuss the $m=0$ case separately below.

\section{Momentum-space eigenstates and quantum resources}
\subsection{IF}
To quantize neutral fermionic fields in momentum space, we utilize the approach of discretized light cone quantization (DLCQ) \cite{Pauli1985-1}.
In IF coordinates, the time evolution is governed by the IF energy operator $P^0$, which we quantize in a periodic box $x^1\in\qty[-\frac{L}{2},\frac{L}{2}]$ with IF momentum and IF energy given by
\begin{equation}\label{eq:IFquantization}
    \begin{split}
    k^1_n=\frac{2\pi n}{L},\ n=-\frac{N}{2},&-\frac{N}{2}+1,\ldots \frac{N}{2}-1;\\
     (k^0_n)^2-(k^1_n)^2&=m^2.
    \end{split}
\end{equation}
In the limit $L\rightarrow\infty$, the momenta $k^1_n$ densely fill the interval $(-\pi/a,\pi/a]$ and can be approximated as continuous. Applying the Fourier transform
\begin{equation}\label{eq:IFfourier}
    c=\int_{-\pi/a}^{\pi/a} \frac{\mathrm{d}k^1}{\sqrt{2\pi}} \hat c_k e^{-ik^\mu x_\mu}=\int_{-\pi/a}^{\pi/a} \frac{\mathrm{d}k^1}{\sqrt{2\pi}} \hat c_k e^{-i(k^0 x^0+k^1x^1)},
\end{equation}
 the momentum-space form of the IF energy operator is written in diagonal form as
% \begin{subequations}\label{eq:IFhamsys}
\begin{equation}
\begin{split}\label{eq:IFbogoham}
     H_{IF}&=\int_{0}^\infty \mathrm{d}k\ 2\omega_{k} \qty(\hat \eta^\dagger_k \hat \eta_k +\hat \eta^\dagger_{-k}\hat \eta_{-k}-1)\\
     \omega_k&=\sqrt{m^2+k^2}
 \end{split}
\end{equation}
 with $\hat\eta_k=\cos\phi_k\hat c_k+i\sin\phi_k\hat c^\dagger_{-k},\ \tan(2\phi_k)=k/m,$
which is a small $ka$ approximation of the Bogoliubov transformation in \eqref{eq:origbogoham}, consistent with the $a\rightarrow0$ continuum limit.

The IF Hamiltonian \eqref{eq:IFbogoham} is block diagonal with blocks labelled by absolute momentum $\vert k\vert$. The ground state of each block is:
\begin{equation}\label{eq:IFgs}
    \ket{\Phi_{k}}=\qty(\cos\phi_k -i\sin{\phi_k}\hat c^\dagger_k\hat c^\dagger_{-k})\ket{0}_k\ket{0}_{-k}%\otimes\ket{0}_{-k}
\end{equation}
where $\ket{0}_{k}$ is the fermionic vacuum state of $\hat c^\dagger_k\hat c_k$. The overall ground state $\ket{\Phi}=\otimes_{k>0}\ket{\Phi_k}$ is the BCS-type ground state
\begin{equation}\label{eq:bogovacuum}
    \ket{\Phi}=\bigotimes_{k>0}\qty(\cos\phi_k \ket{0}_k\ket{0}_{-k}-i\sin{\phi_k}\ket{1}_k\ket{1}_{-k})
\end{equation}
and excited states are:
\begin{equation}\label{eq:IFeigs}
\ket{k'}=\hat\eta^\dagger_{k'}\ket{\Phi}=\ket{1}_{k}\ket{0}_{-k'}\bigotimes_{\substack{{k>0}\\{k\neq k'}}}\ket{\Phi_k}
\end{equation}
with energies $E_k=2\sqrt{m^2+k^2}$~\cite{Mussardo2020,Mbeng2020,Fradkin2021}.

\subsection{LF}
We will now consider LF momentum-space representation of the LF energy operator \eqref{eq:LFpminus}. Quantizing in a box $x^-\in\qty[-\frac{L}{2},\frac{L}{2}]$ with a periodic boundary condition \footnote{We note that the IF and LF boundary conditions are distinct and thus a particular value of $k^+_n$ does not necessarily translate to an allowed value of $k^1_n$. However, because we are assuming the the thermodynamic limit $N\rightarrow\infty$ and treating $k^1$ and $k^+$ as continuous, we can safely neglect this nuance.}, the LF momentum $k^+$ and LF energy $k^-$ are given by \cite{Pauli1985-1} 
\begin{equation}\label{eq:LFquantization}
    k_n^+a=\frac{2\pi n}{N},\ n=1,2,\ldots,N;\quad k^+_nk^-_n=m^2
\end{equation}
Again assuming $L\rightarrow\infty$, the LF Fourier transform is 
\begin{equation}\label{eq:LFfourier}
\begin{split}
        \psip
        &=\int_{0}^{2\pi/a} \frac{\mathrm{d}k^+}{\sqrt{2\pi}} \check{c}_{k^+} e^{-i(k^+x^-+k^-x^+)/2},
\end{split}
\end{equation}
and we derive from \eqref{eq:lffreefermievol}
\begin{equation}
    \psim=\int_{0}^{2\pi/a} \frac{\mathrm{d}k^+}{\sqrt{2\pi}}\frac{m}{k^+} \check{c}_{k^+} e^{-i(k^+x^-+k^-x^+)/2},
\end{equation}
where $\check{c}$ denotes that the operator is acting in LF momentum space.

Substituting in to the LF energy operator \eqref{eq:LFpminus} and setting $x^+=0$, we then find the momentum space LF Hamiltonian
\begin{equation}\label{eq:LFkham}
    H_{LF}=\int_{0}^{\infty}\mathrm{d}k^+\ \frac{m^2}{2k^+}\check{c}_{k^+}^\dagger \check{c}_{k^+}.
\end{equation}
Hence \eqref{eq:LFkham} is diagonalized in LF momentum space. The eigenstates of the LF momentum-space Hamiltonian \eqref{eq:LFkham} are:
\begin{equation}\label{eq:LFeigs}
    \ket{k^+}=\check{c}^\dagger_{k^+}\ket{0}_{k^+}
\end{equation}
with energies $E_{k^+}=m^2/(2k^+)$. The form of \eqref{eq:IFgs} shows that while the IF eigenstates exhibit entanglement between momentum $+k$ and momentum $-k$, the LF eigenstates \eqref{eq:LFeigs} are entirely separable in both LF and IF momentum-space.

\subsection{Magic}
In addition to entanglement, another resource for quantum computing is the non-stabilizerness or ``magic'', which is associated with the level of non-stabilizerness or contextuality~\cite{Spekkens2008,Bravyi2005,Gottesman1999,Veitch2014,Bravyi2016,Chitambar2019,Liu2022,Beverland2020}. To quantify the non-stabilizerness of the IF and LF eigenstates, we utilize the framework of stabilizer R\'{e}nyi entropy (SRE) \cite{Leone2022,Haug2023,Turkeshi2025,Dora2024}. To evaluate resources in momentum-space, we follow \cite{Dora2024} and define momentum-space qubits through the inverse Jordan-Wigner transformation \footnote{To construct the Jordan-Wigner string operator $\exp\qty(i\pi \sum_{k'<k}\hat c^\dagger_{k'}\hat c_{k'})$ in IF momentum-space, we again follow \cite{Dora2024} and order the IF momenta in a sequence of opposite pairs $-k_1,k_1,-k_2,k_2,-k_3,k_3,\cdots$ where $k_i>0$ but the adjacent pairs need not have any particular order.}
\begin{equation}\label{eq:IFqubitization}
\begin{split}
        \frac{1}{2}\qty(\hat{X}_k+i\hat{Y}_k)&=\hat c^\dagger_{k}\exp\qty(i\pi \sum_{k'<k}\hat c^\dagger_{k'}\hat c_{k'})\\
        \hat{Z}_k&=2\hat c_k^\dagger \hat c_k-1.
\end{split}
\end{equation}
 The SRE of the ground state $\ket{\Phi}$ is then defined as \cite{Leone2022}
\begin{equation}
    M_q\equiv\frac{1}{1-q}\ln\qty(\frac{1}{4^N}\sum_{P\in P^{(N)}}\abs{\mel{\Phi}{P}{\Phi}}^{2q})
\end{equation}
where $P^{(N)}=\{I,X,Y,Z\}^{\otimes N}$ is the set of all $4^N$ possible momentum-space Pauli strings.

Because the LF ground state is separable in LF-momentum-space, it is a stabilizer state with $M^{LF}_2=0$. The IF ground state \eqref{eq:bogovacuum} factorizes by absolute momentum. Taking $\ket{0}_k$ to be the $-1$ eigenstate of $\hat Z_k$ and $\ket{1}_k$ to be the $+1$ eigenstate, the non-zero ground state expectation values in each of the $N/2$ 4-dimensional subspaces are then
\begin{gather}
    \mel{\Phi_k}{\hat{I}_k\hat{I}_{-k}}{\Phi_k}=\mel{\Phi_k}{\hat{Z}_k\hat Z_{-k}}{\Phi_k}=1\nonumber\\    
    \mel{\Phi_k}{\hat{I}_k\hat{Z}_{-k}}{\Phi_k}=\mel{\Phi_k}{\hat{Z}_k\hat{I}_{-k}}{\Phi_k}=-\cos(2\phi_k)\label{eq:revJW}\\
    \mel{\Phi_k}{\hat{X}_k\hat{Y}_{-k}}{\Phi_k}=\mel{\Phi_k}{\hat{Y}_k\hat{X}_{-k}}{\Phi_k}=\sin(2\phi_k),\nonumber
\end{gather}
with the other 10 expectation values being zero \cite{Dora2024}. The ``magic'' $M_2$ is then
\begin{equation}\label{eq:ifmagic}
\begin{split}
    M^{IF}_2&=-\ln\qty(\prod_{k>0}\frac{1+\cos^4(2\phi_k)+\sin^4(2\phi_k)}{2})\\
    &=-\sum_{k>0}\ln\qty(1-\qty(\frac{km}{k^2+m^2})^2).
\end{split}
\end{equation}
Because for $m>0$, the $m=k$ term (corresponding to $\phi_k=\pi/4$) will always contribute $\ln(4/3)\approx0.288$, we can see that for all $m>0$ the IF ground state will require more magic to prepare than the LF ground state.

\section{The phase transition}
We will now consider the behavior at the quantum critical point $\lambda=1$, corresponding to a massless free fermion in the field representation.
The IF Hamiltonian \eqref{eq:IFbogoham} remains well behaved at $\lambda=1$, leaving us with
\begin{equation}\label{eq:IFmasslessham}
    H_\mathrm{IF}^{(m=0)}=\int_{0}^\infty \mathrm{d}k\ 2k \qty(\hat \eta^\dagger_k \hat \eta_k +\hat \eta^\dagger_{-k}\hat \eta_{-k}-1)
\end{equation}
where now $\hat\eta_k=\frac{1}{\sqrt{2}}(\hat{c}_k-i\hat{c}^\dagger_{-k})$. The ground state then becomes the maximally entangled Bell state
\begin{equation}\label{eq:masslessIFGS}
    \ket{\Phi}=\bigotimes_{k>0}\qty(\frac{1}{\sqrt{2}}\ket{0}_k\ket{0}_{-k}-\frac{i}{\sqrt{2}}\ket{1}_k\ket{1}_{-k}).
\end{equation}
Because each $\ket{\Phi_k}$ in \eqref{eq:masslessIFGS} is now a stabilizer in the momentum-space qubitization \eqref{eq:IFqubitization}, the IF magic \eqref{eq:ifmagic} now goes to zero.

A consequence of \eqref{eq:lffreefermievol} is that in the massless case, the ``good'' fermion $\Psiplus$ becomes LF time-invariant and the ``bad'' fermion $\Psiminus$ becomes LF space-invariant, requiring particular care when quantizing these variables.  For our purposes, we follow the approach of \cite{Hornbostel1988} and quantize $P^-$ independently of $k^+$ in the massless case to allow for a direct comparison with the IF energy operator $P^0$; however, we note that the proper LF quantization of massless fields and the connection to zero modes remains an area of active research \cite{Martinovic2015,Martinovic2016}.
From the LF dispersion relation \eqref{eq:LFquantization}, we can see that $k^+$ in the $m=0$ case will no longer be a good variable with which to quantize the LF energy operator $P^-$ as $k^-=0$ for any $k^+>0$ \cite{Martinovic2015,Martinovic2016}.  This leads us, following the approach of \cite{Hornbostel1988}, to instead quantize in terms of $k^-$, resulting in a form of \eqref{eq:LFkham} as
\begin{equation}
    H_{LF}=\frac{1}{2}\int_0^\infty \mathrm{d} k^-\ k^-\check{c}_{k^-}^\dagger\check{c}_{k^-}
\end{equation}
We then note that when $m=0$, \eqref{eq:IFquantization} implies $k^0=\vert k^1\vert$, resulting in two potential cases for our LF quantized energy and momenta:
\begin{equation}
    \begin{split}
        k^1>0:\quad&k^+=2k^1,\ k^-=0\\
        k^1<0:\quad&k^+=0,\ k^-=-2k^1.
    \end{split}
\end{equation}
This leads us to then write
\begin{equation}
    H^{(m=0)}_{LF}=\int_{0}^\infty \mathrm{d}k\  2k\  \hat{c}^\dagger_{-2k} \hat{c}_{-2k},
\end{equation}
which is equivalent to the diagonalized massless IF Hamiltonian \eqref{eq:IFmasslessham} with only negative even momenta included, and the massless LF eigenstates will thus be of the form 
\begin{equation}\label{eq:masslessLFGS}
    \ket{k^-}=\hat{c}^\dagger_{-2k}\ket{0}.
\end{equation}
Thus, even while LF and IF quantization coincide at the quantum critical point, the LF ground state \eqref{eq:masslessLFGS} remains distinct from the IF ground state \eqref{eq:masslessIFGS}, and remains entirely separable in momentum-space while the IF ground state is maximally entangled.

\section{Discussion}
In this paper, we used the transverse field Ising model (TFIM) to explore quantum resources in IF and LF coordinates. We derived the Hamiltonians in both LF and IF momentum space, and show that the LF Hamiltonian is diagonal without the Bogoliubov transformation required in IF. The eigenstates of the LF Hamiltonian are separable in LF momentum space, while the eigenstates of the IF Hamiltonian have pairwise entanglement between modes with momentum $\pm k$ in IF momentum space.  We quantified the non-stabilizerness, or magic, of the ground states using momentum-space stabilizer R\`enyi entropy.  At the quantum critical point $\lambda=1$ (corresponding to a massless free fermion) the IF states are also stabilizer states, though still exhibiting entanglement between positive and negative momenta.

The simplicity of the LF ground state has been known for some time and is an active area of research \cite{Mannheim2019Vacuum, Collins2018,Ullrich2006,Coester1994,Tsujimaru1998}, we see it emerge here simply from the form of LF quantization. The entanglement exhibited by IF ground states emerges from the Kramers-Wannier duality between a particle with momentum $+k$ and a hole with momentum $-k$, which results in a block-diagonal structure in momentum-space. But the form of the LF Hamiltonian $P^-=P^0-P^1$ does not permit such a coupling, breaking the duality and resulting in a LF Hamiltonian which is diagonal in LF momentum without the need for a Bogoliubov transformation. While the LF has been actively studied for high energy and nuclear physics for more than 75 years, this work represents the first discussion of quantum resources on the light front. Extension of the current study to interacting theories, higher dimensions, and consideration of the effects of renormalization are all promising future directions.

\begin{acknowledgments}
We wish to thank Kamil Serafin, Carter Gustin, Michael Kreschuk, Feng Qian, James Vary, and Natalie Klco for useful advice and discussion.  This work was supported by the United States Department of Energy under the Office of Science Quantum Information Science Research Centers for the \textbf{Q}uantum \textbf{S}ystems \textbf{A}ccelerator (QSA) program (7568717) and under the Office of Nuclear Physics Quantum Horizons program for the \textbf{Nu}clei and \textbf{Ha}drons with \textbf{Q}uantum Computers (NuHaQ) project (DE-SC0023707).
\end{acknowledgments}

\appendix*
\section{Appendix}\label{app:isingdiagonalization}
\setcounter{equation}{0}
In this appendix, we more thoroughly derive the diagonalization of the transverse-field Ising model Hamiltonian \eqref{eq:origfermiham} in (IF) momentum space. To work in momentum-space, we utilize the Fourier transforms
\begin{equation}\label{eq:appfourier}
    \hat{c}_k=\frac{1}{\sqrt{N}}\sum_{j=1}^N e^{-ikja}c_j,\quad \hat{c}^\dagger_k=\frac{1}{\sqrt{N}}\sum_{j=1}^N e^{ikja}c^\dagger_j.
\end{equation}
To understand which values of momentum to use, we note that our fermionic boundary condition requires $c_N=c_1$ and thus $\sum_{k}e^{ik(N+1)a}=\sum_{k}e^{ika}$ or $e^{ikNa}=1$ for all $k$. Thus, the $N$ allowable values of $k$ are 
\begin{equation}
    k\in\qty{\frac{2\pi n}{L},\quad n=-\frac{N}{2},-\frac{N}{2}+1,\ldots,\frac{N}{2}-1}.
\end{equation}
For $L\rightarrow\infty$, we can see that the spacing between the allowable momenta vanishes, allowing us to treat the momentum as continuous over $[-\pi/a,\pi/a]$.

We then utilize the identity 
\begin{equation}
    \lim_{N\rightarrow\infty}\sum_{j=1}^N e^{2\pi i j(n-m)/N}=\delta_{n,m},
\end{equation}
which allows us to write the Hamiltonian \eqref{eq:origfermiham} in momentum space as 
\begin{align}
    H_{IF}=\sum_k \bigg[&2(\lambda-\cos(ka))\hat{c}_k^\dagger\hat{c}_{k}-(\lambda-e^{-ika})\nonumber \\
    &-e^{ika}(\hat{c}^\dagger_k\hat{c}^\dagger_{-k}+\hat{c}_k\hat{c}_{-k})\bigg]\nonumber \\
\begin{split}
    =\sum_{k>0} \bigg[&2(\lambda-\cos(ka))(\hat{c}_k^\dagger\hat{c}_{k}+\hat{c}_{-k}^\dagger\hat{c}_{-k}-1)\\
    &-2i\sin(ka)(\hat{c}^\dagger_k\hat{c}^\dagger_{-k}+\hat{c}_k\hat{c}_{-k})\bigg].
\end{split}\label{eq:appkham}
\end{align}

We observe that the momentum-space Hamiltonian \eqref{eq:appkham} is block-diagonal in momentum space, with each block $H_k$ corresponding to a single absolute momentum $k$. Each block is made up of four terms: a number operator $\hat{c}_k^\dagger \hat{c}_k$ for particles with momentum $+k$; a number operator $\hat{c}_{-k}^\dagger \hat{c}_{-k}$ for particles with momentum $-k$; a creation operator $\hat{c}_k^\dagger \hat{c}_{-k}^\dagger$ which creates a pair of particles, one with momentum $+k$ and one with momentum $-k$; and the corresponding annihilation operator $\hat{c}_k \hat{c}_{-k}$. 

To diagonalize the blocks, we introduce the Bogoliubov transformation
\begin{equation}
    \hat\gamma_k=\cos{\theta_k}\hat{c}_k+i\sin{\theta_k}\hat{c}^\dagger_{-k}.
\end{equation}
Subsititing in to \eqref{eq:appkham}, we find 
\begin{widetext}
\begin{equation}
    \begin{split}
    H_k=&2\big((\lambda-\cos(ka))\cos{2\theta_k}+\sin(ka)\sin{2\theta_k}\big)\qty(\hat{\gamma}^\dagger_k\hat{\gamma}_k+\hat{\gamma}^\dagger_{-k}\hat{\gamma}_{-k}-1)\\
    &+2i\big(-(\lambda-\cos(ka))\sin{2\theta_k}+\sin(ka)\cos{2\theta_k}\big)\qty(\hat\gamma^\dagger_k\hat\gamma^\dagger_{-k}+\hat\gamma_k\hat\gamma_{-k}).
    \end{split}
\end{equation}
\end{widetext}

We can thus see that $H_k$ will be diagonalized when
\begin{equation}
-(\lambda-\cos(ka))\sin{2\theta_k}+\sin(ka)\cos{2\theta_k}=0,
\end{equation}
giving us the constraint equation for $\theta_k$
\begin{equation}
    \tan2\theta_k=\frac{\sin(ka)}{\lambda-\cos(ka)}.
\end{equation}
We can then write 
\begin{equation}
    \begin{split}
    \cos{2\theta_k}&=\frac{\lambda-\cos(ka)}{\sqrt{\lambda^2-2\lambda \cos(ka)+1}},\\
     \sin{2\theta_k}&=\frac{\sin(ka)}{\sqrt{\lambda^2-2\lambda \cos(ka)+1}}
    \end{split}
\end{equation}
and thus
\begin{widetext}
\begin{equation}
    (\lambda-\cos(ka))\cos{2\theta_k}+\sin(ka)\sin{2\theta_k}=\sqrt{\lambda^2-2\lambda\cos(ka)+1}\equiv\omega_k.
\end{equation}
\end{widetext}
We then arrive at 
\begin{equation}
    H_{IF}=\sum_{k}2\omega_k\qty(\hat{\gamma}^\dagger_k\hat{\gamma}_k+\hat{\gamma}^\dagger_{-k}\hat{\gamma}_{-k}-1).
\end{equation}

\bibliography{IsingLFPaper.bib}
\end{document}